\documentstyle[preprint,tighten,aps]{revtex}
\begin{document}
\preprint{TTP95-14, hep-ph/9507420}
\draft
\title{
Electroweak Radiative Corrections To Polarized M{\o}ller Scattering
Asymmetries}
\author{Andrzej Czarnecki}
\address{Institut f\"ur Theoretische Teilchenphysik,
Universit\"at Karlsruhe,\\
D-76128 Karlsruhe, Germany}
\author{William J.~Marciano}
\address{Physics Department,
Brookhaven National Laboratory,\\
Upton, New York 11973\\
and\\
Institute for Nuclear Theory, Box 351550, \\University of Washington,
Seattle, WA 98195-1550}
\maketitle

\begin{abstract}
One loop electroweak radiative corrections to left-right parity violating
M\o ller scattering ($e^-e^-\to e^-e^-$) asymmetries are presented. They
reduce the standard model (tree level) prediction by 40$\pm 3$ \% where the
main shift and uncertainty stem from hadronic
vacuum polarization loops.  A similar reduction also occurs for the
electron-electron atomic parity violating interaction.
That effect can be attributed to an increase of $\sin^2\theta_W(q^2)$ by
$3\%$ in running from $q^2=m_Z^2$ to 0.  The sensitivity of the asymmetry to
``new physics'' is also discussed.
\end{abstract}

\narrowtext
\section{Introduction}
The chiral structure of the standard $SU(2)_L \times U(1)_Y$ model implies
a predictable degree of parity violation in all physical processes, ranging
from low energy atomic phenomena to high energy $Z$ boson production
asymmetries.  Precision experimental studies of those predictions test the
standard model at the tree and quantum loop level.  A deviation from
expectations would point to ``new physics''.

One interesting class of parity violation measurements involves the
scattering of longitudinally polarized (left or right-handed) electrons on
an unpolarized target.  The left-right scattering asymmetry
\begin{equation}
A_{LR} \equiv
{{\rm d}\sigma_L- {\rm d}\sigma_R
\over
 {\rm d}\sigma_L+ {\rm d}\sigma_R
}
\end{equation}
is manifestly parity violating and measures the interference between
electromagnetic and weak neutral current amplitudes. A classic example is
the now famous SLAC asymmetry measurement for deep-inelastic polarized
$e-D$ scattering \cite{SLACeD}. That study confirmed the standard model's
weak neutral current structure and provided a good determination of the
weak mixing angle, $\sin^2\theta_W$ (to about $\pm$10\%). One could
envision pushing such asymmetry measurements to much higher levels of
precision. Indeed, a later measurement of elastic polarized $e-C$ scattering
\cite{eCscat} indicated that systematic uncertainties as small as $10^{-8}$
were achievable in asymmetry experiments.

Given the possibility of very high precision asymmetry measurements using
present day facilities and technology, it is interesting to investigate
what one can learn from such experiments. In that spirit, we consider here
the case of polarized M\o ller scattering $e^-e^-\to e^-e^-$.  Our primary
focus will be on the use of a very intense highly polarized ($P_e > 0.8$)
electron beam in fixed target unpolarized electron scattering.

The tree level prediction for that asymmetry was examined a number of years
ago \cite{tree}.  The interference between electromagnetic and weak neutral
current amplitudes in fig.~1 gives rise to the standard model prediction
\begin{eqnarray}
\lefteqn{A_{LR}(e^-e^-\to e^-e^-)}
\nonumber\\
&&  \qquad = {G_\mu Q^2 \over \sqrt{2} \pi \alpha}
{1-y \over 1+y^4+(1-y)^4 } \left(1-4\sin^2 \theta_W\right)
\label{eq:treeA}
\end{eqnarray}
where
\begin{eqnarray}
G_\mu &=& 1.16639(1)\times 10^{-5} \,{\rm GeV}^{-2} \nonumber\\
\alpha^{-1} &=& 137.036 \nonumber\\
Q^2 &=& -q^2 \equiv y(p^\prime+p)^2 = y(2m_e^2+2m_e E_{\rm beam})_{\rm
fixed\,\, target}
\nonumber\\
q^2&=& (p^\prime-p)^2
\label{eq:notation}
\end{eqnarray}
and the weak mixing angle is roughly $\sin^2\theta_W\approx 0.23$. In
that expression, terms of order $m_e/E_{\rm beam}$ and $m_e/Q$ have
been dropped, since we assume $m_e^2\ll Q^2 \ll m_Z^2$.

For fixed target experiments, the asymmetry in (\ref{eq:treeA}) is very
small because of the tiny $G_\mu Q^2$ factor and (to a lesser extent) the
$1-4\sin^2\theta_W$ suppression factor. Employing a $Z$ pole value,
$\sin^2\theta_W=0.2314$, and choosing $y=1/2$ where the asymmetry is
maximal, one finds (for $100\%$ beam polarization, $P_e=1$) the tree level
prediction
\begin{equation}
A_{LR}(e^-e^-\to e^-e^-) \approx 6\times 10^{-9}
(E_{\rm beam}/1 {\rm GeV})
\end{equation}
That small an asymmetry may at first sight appear impossible to measure. An
experimental group has, however, taken up the challenge and studied the
possibility of such a measurement \cite{expMoller}. They envision using the
SLAC 50 GeV beam (such that $A_{LR}^{\rm tree}\approx 3\times
10^{-7}$) and operating
with very high, well monitored polarization $\left|P_e\right| \gtrsim 0.8$.
They estimate
that using a thick hydrogen target, a statistical precision of $\pm
10^{-8}$ in $A_{LR}$ is achievable in a 3 month run.  That corresponds to
an accuracy of $\pm 3\%$ of the standard model tree level prediction
and a determination of $\sin^2\theta_W$ to $\pm 0.0006$.
Keeping systematic uncertainties at or
below that level is difficult, but its
technical feasibility has been experimentally demonstrated.
Indeed, the experimental feasability study suggests that a measurement of
$A_{LR}$ with a total error of $\pm 1.4 \times 10^{-8}$ is possible.

The number of scattering events required for a $10^{-8}$ statistical
accuracy is very large, $\sim 10^{16}$.  However, such a large data set
requirement is not so daunting when one considers the gigantic cross-section
in M\o ller scattering at low $Q^2$. (A realistic experiment at SLAC would
have $\langle Q^2 \rangle \approx 0.02 \,{\rm GeV}^2$.)

A measurement of $\Delta A_{LR}$ to $\pm 1.4\times 10^{-8}$
is only useful if one
knows the standard model prediction to that level of certainty. Such
precision requires the inclusion of quantum loop effects.  Indeed, because
the tree level prediction is suppressed by $1-4\sin^2\theta_W$, one
anticipates that the relative size of one loop contributions without such a
suppression factor will be quite big and that indeed turns out to be the
case. In section \ref{sec:RadCor},
we present the complete one loop radiative corrections
to $A_{LR}$ and show that they reduce the standard model prediction by
about 40\%. That reduction results mainly from $\gamma-Z$ mixing via
hadronic vacuum polarization effects. Hadronic loops necessarily entail
theoretical uncertainty. However, we show that the uncertainty is
conservatively at the $\pm 10^{-8}$ level in the experiment under
discussion and thus well matched to envisioned experimental errors. We
describe how the theoretical uncertainty could be further reduced by future
studies.  We also show how the reduction in $A_{LR}$ can be viewed as the
running of $\sin^2\theta_W(q^2)$ as $q^2$ varies from $m_Z^2$ to
$|q^2|\approx 0.02$ ${\rm GeV}^2$
which is of relevance for M\o ller scattering in the
planned fixed target experiment.

As a byproduct of our study, we also show that the electron-electron
parity violating neutral current interaction is similarly reduced by
about 40\% with respect to tree level expectations.

Given the possibility of measuring $\Delta A_{LR}$ to $\pm 1.4\times
10^{-8}$, one can also ask what ``new physics'' would be probed?
Also, how does such a measurement compare with other precision
studies, such as Atomic Parity Violation which has already reached the
1-2\% level and where further improvement is anticipated?  To
illustrate the utility of polarized $e^-e^-$ scattering, we examine in
Section \ref{sec:NewPhys} several ``new physics'' scenarios such as
effects of $Z^\prime$ bosons, S, T, U, V, W and X loop effects, and
constraints on an anomalous electron anapole moment.  The potential of
a $\pm 1.4\times 10^{-8}$ measurement of $A_{LR}$ is compared with
various other precision electroweak experiments, particularly atomic
parity violation.

In Section \ref{sec:Sum}, we summarize our conclusions and comment on
possible future expectations.

\section{One loop electroweak radiative corrections}
\label{sec:RadCor}
Specification of the one loop radiative corrections to $A_{LR} (e^-e^-)$
requires that we properly define the renormalized parameters that are used
in the tree level expression.  Our prescription is fairly conventional. We
choose $G_\mu$ defined by the muon lifetime formula~\cite{beh56,kin59}
\begin{eqnarray}
\tau^{-1}_\mu &=& {G_\mu^2 m_\mu^5\over 192\pi^3}
f\left( {m_e^2\over m_\mu^2} \right) \left( 1+ {3\over 5}{m_\mu^2\over
m_W^2} \right)
\nonumber\\
&&\qquad \times
\left[ 1+ {\alpha(m_\mu)\over 2\pi}\left( {25\over 4}-\pi^2\right)\right]
\nonumber\\
f(x) &\equiv & 1-8x+8x^3-x^4-12x^2\ln x,
\nonumber\\
\alpha(m_\mu) &\approx& 1/136
\end{eqnarray}
That definition leads to the value of $G_\mu$ in (\ref{eq:notation}). Of
course, many of the loop corrections to muon decay have been absorbed into
$G_\mu$. Those corrections are needed when we express neutral current
amplitudes in terms of $G_\mu$ and will give rise to part of the radiative
corrections to $A_{LR}$.  Fortunately, those effects are known from
previous studies \cite{frame,neutr,atom}.

The fine structure constant $\alpha$ in (\ref{eq:treeA}) is defined by
Thomson scattering at $q^2=0$ and found to have the value in
(\ref{eq:notation}). That quantity is a holdover from atomic physics
studies and not always appropriate as a weak loop expansion parameter. For
that reason, we prefer to employ $\alpha(m_Z)$
\begin{eqnarray}
\alpha^{-1}(m_Z)=127.9\pm 0.1
\end{eqnarray}
defined by $\overline{\rm MS}$ (modified minimal subtraction)
at $\mu=m_Z$ in short distance dominated loop corrections.  By that
judicious choice, we avoid inducing 2 loop effects that would be $\sim$7\%
of the one loop corrections.  Note, however, that some of the most
important loop corrections (in particular $\gamma Z$ mixing loops) are
better (and more appropriately) parametrized by $\alpha$ \cite{MS81}.

The renormalized weak mixing angle will be defined by $\overline{\rm MS}$
at scale $\mu=m_Z$, $\sin^2\theta_W(m_Z)_{\overline{\rm MS}}$. The use of
that scheme simplifies the form of the radiative corrections.  For readers
more comfortable with $\sin^2\theta_W^{\rm eff}$ used in LEP and SLC
asymmetries, there is a simple numerical translation \cite{GS94}
\begin{eqnarray}
\sin^2\theta_W(m_Z)_{\overline{\rm MS}}
=
\sin^2\theta_W^{\rm eff}-0.0003
\end{eqnarray}
The analytic form of the radiative corrections in that translation is
extremely complicated and will not be given here.

For input, we use
\begin{eqnarray}
\sin^2\theta_W(m_Z)_{\overline{\rm MS}}
= 0.2314
\end{eqnarray}
which is consistent with $Z$ pole measurements as well as the indirect
determinations that use $\alpha$, $G_\mu$ and $m_Z=91.190$ GeV along with
\begin{eqnarray}
m_t(m_t)_{\overline{\rm MS}}
&\equiv& m_t = 170 \,{\rm GeV}
\nonumber \\
m_H &=& {\rm (Higgs\,\, Mass)} = 200 \,{\rm GeV}
\end{eqnarray}
That input requires for standard model consistency, $m_W=80.39$ GeV, a
value we also adhere to.

Given the above renormalization prescription, we can now unambiguously
write down the one loop radiative corrections to $A_{LR}
(e^-e^-)$. Some parts can be obtained from existing calculations while
others require a new study. In total, we find (\ref{eq:treeA}) is
modified as follows
\begin{eqnarray}
A_{LR} (e^-e^-) &=&
{\rho G_\mu Q^2 \over \sqrt{2} \pi \alpha}
{1-y \over 1+y^4+(1-y)^4 }
\nonumber \\
   \times  \left\{ \rule{0mm}{5mm} \right.
 1&-&4\kappa(0)\sin^2 \theta_W (m_Z)_{\overline{\rm MS}}
       +{\alpha(m_Z)\over 4\pi s^2}
\nonumber\\
    &-& {3\alpha(m_Z)\over 32\pi s^2c^2}(1-4s^2)[1+(1-4s^2)^2]
\nonumber\\
    &+&F_1(y,Q^2) +F_2(y,Q^2)  \left. \rule{0mm}{5mm}  \right\}
\label{eq:corr}
\end{eqnarray}
where
\begin{eqnarray}
s&\equiv& \sin \theta_W (m_Z)_{\overline{\rm MS}}
\nonumber\\
c&\equiv& \cos \theta_W (m_Z)_{\overline{\rm MS}}
\end{eqnarray}
The quantity $\rho = 1+{\cal O}(\alpha)$ comes about because we have chosen
to normalize the weak neutral current amplitude in terms of the muon decay
constant $G_\mu$. From earlier work \cite{neutr},
one finds that the renormalization of
$G_\mu$ combined with vertex and self-energy renormalizations of the $Z$
amplitude gives
\begin{eqnarray}
\rho &=& 1+{\alpha(m_Z)\over 4\pi} \left\{
{3\over 4s^4}\ln c^2- {7\over 4s^2} + {3\over 4s^2}
{m_t^2(m_t)_{\overline{\rm MS}} \over m_W^2} \right.
\nonumber\\
 &&
\left.\qquad\qquad\qquad
 +{3\over 4} {\xi \over s^2}
\left({\ln(c^2/\xi)\over c^2-\xi} + {1\over c^2} { \ln\xi\over 1-\xi}
\right)\right\}
\nonumber\\
\xi&=& m_H^2/m_Z^2
\end{eqnarray}
Numerically, for a Higgs mass, $m_H\,=\,200$ GeV, and top mass
$m_t(m_t)_{\overline{\rm MS}}$ = 170 GeV, one finds
\begin{eqnarray}
\rho=1.00122
\end{eqnarray}
The smallness of that correction is due to accidental cancellations.

The most important loop corrections are embodied in $\kappa(0) = 1+
{\cal O}(\alpha)$. They come from $\gamma-Z$ mixing and the anapole moment
diagrams illustrated in fig.~2. They are normalized at $Q^2=0$.  Effects
due to $Q^2\neq0$ are absorbed in $F_2(y,Q^2)$ which will be discussed
later.  Evaluated in a free field framework (i.e.~ignoring strong
interactions for the moment)
\begin{eqnarray}
\kappa(0)&=&1-{\alpha\over 2\pi s^2}\left\{
{1\over 3}\sum_f (T_{3f}Q_f -2s^2Q_f^2) \ln{m_f^2\over m_Z^2} \right.
\nonumber\\ &&\left.
-\left( {7\over 2}c^2+{1\over 12}\right) \ln c^2 +
\left( {7\over 9}-{s^2\over 3}\right) \right\}
\label{eq:kappa0}
\end{eqnarray}
where $T_{3f} =\pm 1/2$ (weak isospin) and $Q_f $= fermion electric charge.
The sum over all fermions (quarks and leptons) with mass $<m_Z$
comes from diagram 2a. (The top quark decouples completely from $\kappa(0)$
because of the specific
definition of $\sin^2\theta_W(m_Z)$ we are using \cite{GS94}.) The
second and third terms stem from diagrams 2b and 2c respectively.

The quark contributions in (\ref{eq:kappa0}) cannot be properly
accounted for perturbatively. Instead, one must use a dispersion
relation to relate those vacuum polarization effects to $e^+e^-\to
{ hadrons}$ data. Such an analysis replaces the quark sum in
(\ref{eq:kappa0}) by \cite{atom,WM93slac}
\begin{eqnarray}
{1\over 3}\sum_{\rm quarks} (T_{3f}Q_f -2s^2Q_f^2) \ln{m_f^2\over m_Z^2}
\to -6.88\pm 0.50
\label{eq:hadsum}
\end{eqnarray}
where the error assigned $\pm 0.50$ is rather conservative.  We
suspect that it would be lowered somewhat by an updated analysis of
$e^+e^-\to {hadrons}$ data.  Such a study may, one day, be
important, since the error in (\ref{eq:hadsum}) will turn out to be
the dominant theoretical uncertainty and close to the projected
experimental error presently attainable.

Numerically evaluating (\ref{eq:kappa0}), one finds
\begin{eqnarray}
\kappa(0) = 1.0301\pm 0.0025
\end{eqnarray}
That correction is very significant. It reduces the predicted $A_{LR}$
by about 38\%. The reason for that sensitivity is the fact that the
quark loop diagrams in fig.~2 are not suppressed by
$1-4s^2$. Alternatively, one
can say that $\kappa(0) \sin^2 \theta_W (m_Z)_{\overline{\rm MS}}$ is
the effective low energy mixing angle appropriate for small $Q^2\sim
0$ rather than $\sin^2 \theta_W (m_Z)_{\overline{\rm MS}}$. The 3\%
increase due to the running of $\sin^2 \theta_W$ gets enhanced because
of the $1-4s^2$ sensitivity.

The next source of one loop corrections comes from the $WW$ and $ZZ$
box diagrams in fig.~3.  The $WW$ box is not suppressed by $1-4s^2$
and gives rise to the $\alpha(m_Z)/ 4\pi s^2$ term in (\ref{eq:corr}).
Taken alone that diagram gives a 4\% enhancement of $A_{LR}$ relative
to the lowest order prediction.  The $ZZ$ box diagrams are suppressed
by $1-4s^2$.  Hence, their contribution, the
$3\alpha(m_Z)(1-4s^2)[1+(1-4s^2)^2]/ 32\pi s^2c^2$ term in
(\ref{eq:corr}) is tiny, ${\cal O}(0.1\%)$.

The next set of loops is illustrated in fig.~4. Together with photonic
corrections to the external legs and vertices in fig.~1 and two photon
exchange diagrams, they give rise to $Q^2$ dependent corrections denoted
by $F_1(y,Q^2)$ in  (\ref{eq:corr}). We find
\begin{eqnarray}
\lefteqn{F_1(y, Q^2)=
-{\alpha\over 4\pi}(1-4s^2)
   \left\{
{22\over 3} \ln{y m_Z^2\over Q^2}
+{85\over 9} +f(y)
   \right\},}
\nonumber \\
\lefteqn{f(y)= -{2\over 3}\ln\left[y(1-y)\right]
+ {1\over (1-y+y^2)^2} \left\{ \rule{0mm}{5mm} \right. }
\nonumber\\ &&
- 2\,\left( 1 - y \right)
  \left( 3 - 3\,y + 4\,{y^3} - 3\,{y^4} \right)\, \ln (1 - y)
\nonumber\\ &&
  - 2\,y\,\left( 1 + 3\,y - 6\,{y^2} + 8\,{y^3}
- 3\,{y^4} \right) \,    \ln (y)
\nonumber\\&&
+ \left( 1 - y \right) \left( 2 - 2\,y - 7\,{y^2}
+ 10\,{y^3} - 8\,{y^4} + 3\,{y^5} \right)  \ln^2 (1 - y)
\nonumber\\&&
 -  y\,\left( 2 - 3\,y - 5\,{y^2} + 8\,{y^3} - 7\,{y^4}
+ 3\,{y^5} \right) \,    \ln^2 (y)
\nonumber\\&&
 + \,\left( 2 - 4\,y + 11\,{y^3} - 13\,{y^4} + 9\,{y^5} -
      3\,{y^6} \right)
\nonumber \\&&\qquad \qquad\qquad\qquad \times
\left[\pi^2-2\ln (1 - y)\,\ln (y)\right]
\left. \rule{0mm}{5mm} \right\}
\label{eq:F1}
\end{eqnarray}
For the maximum asymmetry, $y=1/2$, one finds
\begin{eqnarray}
f\left({1\over 2}\right)=
{17\over 12}\pi^2 +{70\over 9}\ln 2-{8\over 3}\ln^2 2\approx 18.09
\end{eqnarray}
The actual evaluation of $F_1$ requires a value of $\sin^2\theta_W$.
Should we use $\sin^2 \theta_W (m_Z)_{\overline{\rm MS}}=0.2314$ or
$\kappa(0) \sin^2 \theta_W (m_Z)_{\overline{\rm MS}}=0.2384$ in
(\ref{eq:F1})? A proper treatment requires a renormalization group
analysis of higher order leading logs. Instead of carrying out that
study, we use the average of those two values and use their spread to
estimate a theoretical uncertainty. In that way, we find for $\langle
Q^2 \rangle =0.02$ ${\rm GeV}^2$
\begin{eqnarray}
F_1(1/2,\,0.02\, {\rm GeV}^2)=-0.0041\pm 0.0010
\end{eqnarray}

The final contribution that we need to consider is the effect of
vacuum polarization in the $\gamma\gamma$ and $\gamma Z$ mixing
self-energies for $Q^2\neq 0$. Because we have chosen to normalize
$\alpha$ and $\kappa(0)$ at zero momentum transfer, there can be a
correction for $Q^2$ non zero.  Fortunately, the residual $Q^2\neq 0$
loop contributions largely cancel out (particularly for $y=1/2$). In
terms of the $\gamma\gamma$ and $\gamma Z$ vacuum polarization
function $\Pi_{\gamma\gamma}$ and $\Pi_{\gamma Z}$, one finds
\begin{eqnarray}
\lefteqn{F_2(y,Q^2)=}
\nonumber\\
&& -4cs\left[
 {1\over 2} \left(
\Pi_{\gamma Z} (-Q^2)
+\Pi_{\gamma Z} \left(-{1-y\over y}Q^2\right)
            \right)
-\Pi_{\gamma Z}(0)\right]
\nonumber\\
&&+\left(1-4s^2\right)
\nonumber\\
&&\qquad\times \left[
 {1\over 2} \left(
\Pi_{\gamma \gamma} (-Q^2)
+\Pi_{\gamma \gamma} \left(-{1-y\over y}Q^2\right)
            \right)
-\Pi_{\gamma \gamma}(0)\right]
\nonumber\\
&&-\left(1-4s^2\right)\left({1\over 2}-y\right)
{1+y(1-y) \over 1-y(1-y)}
\nonumber\\
&&\qquad\times
\left[
\Pi_{\gamma \gamma} \left(-{1-y\over y}Q^2\right)
-\Pi_{\gamma \gamma} (-Q^2)
\right]
\nonumber\\
\end{eqnarray}
For $y=1/2$, the last piece vanishes and lepton loops completely cancel.
One finds
\begin{eqnarray}
\lefteqn{F_2(y=1/2,Q^2)=}
\nonumber\\ &&\qquad\qquad
-4cs\left.\left[
\Pi_{\gamma Z} (-Q^2)
-\Pi_{\gamma Z}(0)\right]\right|_{\sin^2\theta_W=1/4}
\label{eq:vacpol}
\end{eqnarray}
where the partial cancellation of hadronic loops is simply accounted
for by evaluating $\Pi_{\gamma Z}$, the vacuum polarization function,
at $\sin^2\theta_W=1/4$.

A proper evaluation of (\ref{eq:vacpol}) requires a study of
$e^+e^-\to {hadrons}$ data via dispersion relations similar to what
went into (\ref{eq:hadsum}). However, for $Q^2$ relatively small, one can
approximate hadronic contributions to $\Pi_{\gamma Z} (-Q^2)
-\Pi_{\gamma Z}(0)$ using a pion loop calculation.  That rough approach
gives
\begin{eqnarray}
F_2(y=1/2,Q^2)_{\rm pions}&\approx& {\alpha \over 4\pi} \left(
{A^3\over 3}\ln{A+1\over A-1} -{2\over 9} -{2\over 3} A^2\right)
\nonumber \\
A&\equiv& \left( 1+{4m_\pi^2\over Q^2}\right)^{1/2}
\end{eqnarray}
For $Q^2\approx 0.025\, {\rm GeV}^2$, the maximum at SLAC, one finds
\begin{eqnarray}
F_2(1/2,\,0.025\, {\rm GeV}^2)\approx 2\times 10^{-5}
\end{eqnarray}
which is negligible.  So, it seems, that for any foreseeable fixed target
effort one can neglect $F_2$.  It is in the noise.  Of course, if $Q^2\gg
m_\pi^2$, a careful evaluation of $F_2(y,Q^2)$ would have to be
undertaken.

The last issue that must be addressed is the effect of bremsstrahlung
on $A_{LR}$.  We have not included that effect because it is dependent
on the kinematic acceptance of a given experiment.  However, we do
expect on general grounds, that bremsstrahlung is relatively
unimportant. Our reasoning is as follows: soft photon effects,
including radiation damping, factorize and cancel in the asymmetry
ratio. Hard bremsstrahlung should also largely cancel, although the
degree of cancellation probably depends on details of the experimental
geometry.  What contribution remains is proportional to ${\alpha\over
\pi}(1-4\kappa(0)\sin^2\theta_W(m_Z)_{\overline{\rm MS}})$ and hence,
likely to be very small. Therefore, neglect of bremsstrahlung, seems
justified at the level of theoretical and experimental uncertainties
we are considering.  Of course, if a specific experiment is carried
out, correcting for bremsstrahlung effects is straightforward and
should be addressed by the experimentalists.

Collecting all of the one loop radiative corrections, one finds for
$y=1/2$ and $Q^2= 0.025\, {\rm GeV}^2$
\begin{eqnarray}
1-4\sin^2\theta_W&\to& 1.00122 \left[ 1 - 4(1.0301\pm 0.0025) (0.2314)
\right.
\nonumber\\
&&\left.
+ 0.0027-0.0001-0.0041\pm 0.0010\right]
\nonumber\\
\end{eqnarray}
or
\begin{eqnarray}
0.0744\to 0.0450\pm 0.0023\pm 0.0010
\end{eqnarray}
That represents a $40\pm 3\%$ reduction in the asymmetry due to
quantum loop effects. The reduction is rather insensitive to $y$ or
$Q^2$ (unless we go to extreme values).  That 40\% reduction also
(roughly) applies to the parity violating electron-electron
interaction of interest in atomic parity violation \cite{BSJ92}.  (In
fact, the reduction there is about 43\%.)  It renders what was already
a tiny effect essentially negligible.

For $Q^2= 0.025\, {\rm GeV}^2$ and $y=1/2$, as envisioned in a
potential SLAC experiment, one finds that the radiative corrections
reduce $A_{LR} (e^-e^-)$ from $2.97\times 10^{-7}$ to $(1.80\pm 0.09
\pm 0.04) \times 10^{-7}$. The theoretical uncertainties in that
result are roughly at the level of present experimental statistical
capabilities.  They are, however, somewhat conservative. One could
imagine that further scrutiny of $e^+e^-\to {hadrons}$ data and
use of the renormalization group to incorporate higher order leading
logs could reduce the theoretical errors by about a factor of 2.
Hence, theory and realistic experimental precision are well matched.

A measurement of $A_{LR} (e^-e^-)$ to $1.4\times 10^{-8}$ may actually
be made easier because of the reduction we have found. Indeed, some
systematic uncertainties which depend on polarization monitoring
uncertainties are proportional to $A_{LR}$ and hence also reduced by
$40\%$.

{}From our results, one sees that a determination of $A_{LR}$ to $\pm
1.4\times 10^{-8}$ measures the standard model radiative corrections
at about the 7 sigma or more level. Those corrections stem mainly from
$\gamma Z$
vacuum polarization effects and can be viewed as the running of
$\sin^2 \theta_W (\mu)_{\overline{\rm MS}}$ from its value 0.2314 at
$\mu=m_Z$ to a 3\% larger value at $\mu=0$. Confirming that loop
prediction of the standard model would certainly
be an important result.  Of course, such sensitivity implies that a
measurement of $A_{LR}$ to $\pm 1.4\times 10^{-8}$ is likely to also
be a good probe of ``new physics''. We, therefore, now describe its
potential for several examples of physics beyond the standard model.

\section{``New physics'' sensitivity}
\label{sec:NewPhys}
Comparison of a precise measurement of $A_{LR}$ with the standard
model prediction can provide a sensitive probe of ``new physics''.  It
requires, of course, a ``new physics'' contribution to the parity
violating $e^-e^- \to e^-e^-$ amplitude.  Also, $A_{LR}$ can indicate
a deviation from the standard model, but cannot specify the source.
Nevertheless, it is instructive to examine various ``new physics''
scenarios and compare their implications for $A_{LR}$ and other
precision measurements.  Here, we consider a few representative
examples.  For each case, we quote the $1\,\sigma$ reach of $A_{LR}$,
assuming a standard model central prediction of $1.8 \times 10^{-7}$
(for $y=1/2$ and $Q^2=0.025$ GeV$^2$) and a total uncertainty
(experimental and theoretical) of $\pm 1.4 \times 10^{-8}$, i.e.~a
$\pm 7.8\%$ confrontation.

\subsection{$Z'$ bosons}
Grand unified theories, such as $SO(10)$ and $E_6$, often predict the
existence of additional neutral gauge bosons, collectively called
$Z'$s. The masses of those particles are not specified, but could
under certain conditions be relatively light, ${\cal O}$(1 TeV), and
nevertheless beyond the reach of current experiments.  For definiteness, we
consider the $E_6$ model \cite{WM93slac} which contains two $Z'$
eigenstates (with $m_{Z_\beta}<m_{Z'_\beta}$)
\begin{eqnarray}
Z_\beta &=& Z_\chi \cos \beta +Z_\psi \sin \beta \nonumber \\
Z'_\beta &=& -Z_\chi \sin \beta +Z_\psi \cos \beta \nonumber \\
&&-{\pi\over 2} \leq \beta \leq {\pi\over 2}
\end{eqnarray}
$E_6$ symmetry specifies the couplings to electrons (up to some
renormalization uncertainties) and one finds that $A_{LR}$  is
increased by
a factor \cite{WM93slac}
\begin{eqnarray}
\lefteqn{1 + 7
\left\{
{m_Z^2\over m^2_{Z_\beta}}
\left( \cos^2\beta + \sqrt{5\over 3} \sin\beta \cos \beta\right)\right.}
\nonumber \\
&&\qquad\qquad\left. +{m_Z^2\over m^2_{Z'_\beta}}
\left( \sin^2\beta - \sqrt{5\over 3} \sin\beta \cos \beta\right)
\right\}
\end{eqnarray}
For an (effective) $SO(10)$ model, $\beta=0$, that expression
simplifies to
\begin{eqnarray}
1 + 7 {m_Z^2\over m^2_{Z_\chi}}
\end{eqnarray}
Hence, at the $1\,\sigma$ level, $m_{Z_\chi} \approx 870 $ GeV is
probed.  That reach is roughly equivalent to a $\pm 1\%$
determination of atomic parity violation in cesium
\cite{Noecker88,London86,WM90}.  It is also comparable to the
discovery reach of an upgraded Tevatron $p\bar p$ collider.

\subsection{Electron Anapole Moment}
The electron matrix element of the electromagnetic current,
$J_\mu^{\rm em}$, can be written as (with $q=p'-p$)
\begin{eqnarray}
\lefteqn{\!\!\!\!\!\!\!
\langle e(p')| J_\mu^{\rm em} | e(p) \rangle = \bar u_e(p')
\Gamma_\mu u_e(p)} \nonumber \\
\Gamma_\mu &=&
 F_1(q^2)\gamma_\mu
+ i F_2(q^2) \sigma_{\mu\nu} q^\nu
- F_3(q^2) \sigma_{\mu\nu} q^\nu\gamma_5
\nonumber\\
&&+F_A(q^2) \left( \gamma_\mu q^2 - 2m_e q_\mu\right) \gamma_5
\label{eq:emcurrent}
\end{eqnarray}
The first three form factors at $q^2=0$ give the electric charge,
anomalous magnetic moment, and electric dipole moment (in units of
$e$).  All three are physical observables.  The parity violating form
factor $F_A ( q^2) $ at $q^2=0$ is called the anapole moment.  It is
not a direct physical observable and suffers from electroweak gauge
ambiguities.  Indeed, in the standard model it is merely a part of the
total loop corrections to a physical quantity and cannot be uniquely
disentangled.  Nevertheless, it is, in principle, possible that some
forms of ``new physics'' contribute to $A_{LR}$  primarily through the
electromagnetic anapole moment.  Alternatively, one can view
constraints on $F_A(0)$ as providing a figure of merit for comparing
different measurements.

The anapole moment interaction in (\ref{eq:emcurrent}) would shift the
$A_{LR}$  prediction by a factor
\begin{eqnarray}
\left( 1 +
{8\sqrt{2} \pi\alpha\over G_\mu (1-4\sin^2\theta_W)}
F_A(0)\right)
\end{eqnarray}
or in units of the $W$ boson mass
\begin{eqnarray}
\left( 1 + 77 m_W^2 F_A(0)\right)
\end{eqnarray}
Therefore, a measurement of $A_{LR}$  to $\pm 7.8\%$ probes
\begin{eqnarray}
F_A(0) = \pm {1\times 10^{-3} \over m_W^2}
\approx \pm (8\times 10^{-18}\,{\rm cm})^2
\end{eqnarray}
That level of sensitivity compares very favorably with other studies
\cite{Kopp}.
It corresponds to atomic parity violation in cesium at about the $\pm
0.3 \%$ level.

\subsection{The $X$ parameter}
If high mass scale ``new physics'' enters through gauge boson
propagators, it is conveniently studied using the Peskin-Takeuchi $S$,
$T$, and $U$ parameters \cite{STU}.  If the scale of the ``new
physics'' is ${\cal O}(m_Z)$, rather than $\gg m_Z$, that formalism
should be extended to $S$, $T$, $U$, $V$, $W$, $X$
\cite{Maks94,Kundu}. The
additional quantities parametrize changes from $Q^2\approx 0$ to
$m_Z^2$ due to ``new physics'' loops.  In that approach, our
$\kappa(0)$ in eq.~(\ref{eq:kappa0}) gets multiplied by \cite{Kundu}
\begin{eqnarray}
(1-0.032 X)
\end{eqnarray}
A measurement of $A_{LR}$  to $\pm 7.8\%$ or $\Delta \sin^2\theta_W$ to
$\pm 0.0011$ then constrains $X$ at the $\pm 0.14$ level.  That is to
be compared with global fits to all existing data \cite{Kundu} which
currently give $X = 0.38\pm 0.59$. So, an $A_{LR}$  measurement could
improve the constraint by a factor of 4 or so.

\subsection{Generic Loops}
If we parametrize ``new physics'' loop contributions to $A_{LR}$  by a
general parity violating 4 Fermi interaction
\begin{eqnarray}
C {\alpha^2\over M^2} \bar e \gamma_\mu \gamma_5 e \bar e \gamma^\mu e
\end{eqnarray}
with $M$ (roughly) the ``new physics'' mass scale, it modifies
$A_{LR}$  by
a factor
\begin{eqnarray}
\left(1+0.05 C{m_W^2\over M^2}\right)
\end{eqnarray}
In theories with $C\approx 1$, we see that a $\pm 7.8 \%$ measurement
of $A_{LR}$  explores the $M\approx m_W$ scale. That is in keeping
with our
finding that the $WW$ box diagram shifts $A_{LR}$  by about $+7\%$. Of
course, there can be enhancements or suppressions in the case of ``new
physics''.  It would be interesting to compute $C/M^2$ in classes of
low mass supersymmetry models.  That exercise is, however, beyond the
scope of this paper.

\section{Conclusion}
\label{sec:Sum}
We have calculated the one loop electroweak radiative corrections to
the parity violating electron-electron interactions and found a rather
substantial  $40\pm
3\%$ reduction of the tree level prediction.  That result further
reduces (the already insignificant) role of the electron-electron
interaction in atomic parity violation and has interesting consequences
for the left-right asymmetry in polarized M{\o}ller scattering.  It is
clear that any future precision measurement of $A_{LR}$ must be
cognizant of those large corrections.  We also showed that an
experimental determination of $A_{LR}$ at the $\pm 7.8\%$ level
provides a useful and competitive probe of ``new physics''.  Used in
conjunction with other precision measurements and direct high energy
probes it may unveil and help to decipher physics beyond the standard
model.

\acknowledgments
A.C. would like to thank Professor W.~Hollik for many discussions and
advice.  A.C.'s research was supported by BMFT 056 KA 93P.
W.J.M. would like to thank the Institute for Nuclear Theory for its
hospitality during the completion of this work.  This manuscript has
been authored under contract number DE-AC02-76CH00016 with the
U.S. Department of Energy. Accordingly, the U.S. Government retains a
non-exclusive, royalty-free license to publish or reproduce the
published form of this contribution, or allow others to do so, for
U.S. Government purposes.


\begin{figure}
\caption{Neutral current amplitudes leading to the asymmetry $A_{LR}$
at tree level.}
\end{figure}

\begin{figure}
\caption{  $\gamma-Z$ mixing diagrams (a-b)  $W$-loop contribution to
the anapole moment (c).}
\end{figure}

\begin{figure}
\caption{  Box diagrams with two heavy bosons.}
\end{figure}

\begin{figure}
\caption{ Boxes containing one photon and $Z$-loop
contribution to the anapole moment.}
\end{figure}

\end{document}